# PERSPECTIVES OF MICRO-PATTERN GASEOUS DETECTOR TECHNOLOGIES FOR FUTURE PHYSICS PROJECTS[1]

Maxim Titov

CEA Saclay, IRFU/SPP, 91191, Gif sur Yvette, France

A centenary after the invention of the basic principle of gas amplification, gaseous detectors - are still the first choice whenever the large area coverage with low material budget is required. Advances in photolithography and micro-processing techniques in the chip industry during the past two decades triggered a major transition in the field of gas detectors from wire structures to Micro-Pattern Gas Detector (MPGD) concepts, revolutionizing cell-size limitations for many gas detector applications. The high radiation resistance and excellent spatial and time resolution make them an invaluable tool to confront future detector challenges at the frontiers of research. The design of the new micro-pattern devices appears suitable for industrial production. In 2008, the RD51 collaboration at CERN has been established to further advance technological developments of MPGDs and associated electronic-readout systems, for applications in basic and applied research. This review provides an overview of the state-of-the-art of the MPGD technologies and summarizes recent activities for the next generation of colliders within the framework of the RD51 collaboration.

1. INTRODUCTION

Future accelerator facilities are an indispensable step towards precision determination of the Standard Model parameters, constraining of the potential new physics contributions, or, alternatively precision New Physics studies, given LHC yield a hint of new physics presence. The quest for Higgs boson discovery with the multipurpose ATLAS and CMS detectors at the LHC, imposed major detector instrumentation work, and pulled most of the technologies to their performance limits. With many fundamental physics questions within the experimental reach (e.g., precision studies of the Higgs boson couplings and other new particles discovered at the LHC) at the future accelerator facilities, a large R&D effort is needed to develop innovative concepts for radiation detection.

Challenges for novel technologies include improved spatial resolution and segmentation, speed and radiation hardness while minimizing power and cost. Advances on all fronts are needed; these include modern solid state vertex and tracking detectors, gaseous detectors, crystals, cryogenic liquids, readout electronics, services (power, cooling, support and material budget), and trigger and data acquisition systems. Applications are directed at the energy, intensity and cosmic scientific frontiers, with distinct challenges, requirements and solutions:

the luminosity upgrade of the Large Hadron Collider (sLHC), the future Linear Collider (ILC or CLIC), the super high luminosity B-factory (the Belle-II), neutrino experiments and direct dark matter searches, ground-based particle astrophysics and space experiments.

Improvements in detector technology often come from capitalizing on industrial progress. Integration advances in microelectronics and mechanics have resulted in novel, ever more complicated instrumentation. For example, bump bonding technology enabled low capacity connections. The possibility of producing micro-structured semi-conductor devices (with a feature size of 10 μm) and corresponding highly integrated readout electronics led to the success of pixel detectors to achieve unprecedented space-point resolution. Industrial advances in photo-lithography, microelectronics and printed circuits have opened the road for production of micro-structured gas amplification devices.

During the evolution, many novel MPGD structures have arisen from the initial ideas, in general using modern photo-lithographic processes on thin insulating supports. Two designs have emerged, because of ease of manufacturing, operational stability and superior performances for tracking applications: the Gas Electron Multiplier (GEM) [1] and Micro-Mesh Gaseous Structure (Micromegas) [2]. By using pitch size of a few hundred microns, an order of magnitude improvement in granularity over wire chambers, both devices exhibit intrinsic high rate capability ($>10^6$ Hz/mm$^2$), excellent spatial and multi-track resolution (~ 30 μm and ~ 500 μm, respectively), and single photo-electron time resolution in the ns range. Recent aging studies revealed that they might be even less vulnerable to the radiation-damage effects, compared to standard silicon micro-strip detectors, if reasonable precautions are taken on the components quality [3]. Within the broad family of MPGDs, more coarse macro-patterned detectors (e.g., thick-GEMs (THGEM) [4,5] or patterned resistive thick GEM devices (RETGEM) [6]) could offer an interesting economic solution for large-area RICH devices: namely good spatial and time resolutions, large gains (single photo-electron sensitivity), relatively low mass and easy construction - thanks to the intrinsic robustness of the PCB electrodes.

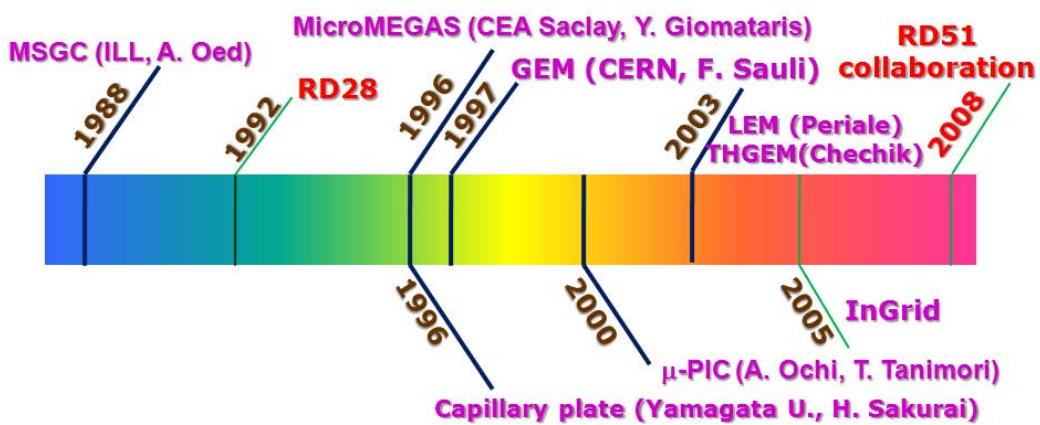

**Fig. 1:** Historical roadmap of the MPGD technology developments.

Coupling of the microelectronics industry and advanced PCB technology has been very important for the development of modern gas detectors with increasingly smaller pitch size. Usually, charge signals in MPGDs are collected on segmented electrodes (strips or pads) and amplified/recorded by external electronics. An elegant solution, that can yield a truly 3D track reconstruction, is the use of the CMOS Application Specific Integrated Circuit (ASIC), assembled directly below the Micromegas or GEM amplification structures [7]-[9]. Modern wafer post-processing technology allows for the integration of a Micromegas grid directly on top of a high-granularity (55 μm pitch) Medipix2 or Timepix chip, thus forming an integrated readout of gaseous detector ("InGrid") [10]. Using this approach, MPGD-based detectors can reach the level of integration, compactness and resolving power typical of solid-state pixel devices.

The historical roadmap for several most advanced MPGD technologies is shown in Fig. 1. The interest in the development and use of the novel micro-pattern gas detectors has led to the establishment in 2008 of the international research collaboration RD51 at CERN [11].

## 2. THE "MPGD" FAMILY

The modern MPGD structures can be grouped in two large families: strip and hole-type structures and micromesh-based detectors. The strip and hole-type structures are: Micro-Strip Gas Chambers (MSGC), GEM, THGEM, RETGEM, Micro-Hole and Strip Plate (MHSP), and Micro-Pixel Gas Chamber (μ-PIC). The micromesh-based structures include: Micromegas, "Bulk" Micromegas, "Microbulk" Micromegas and "InGrid".

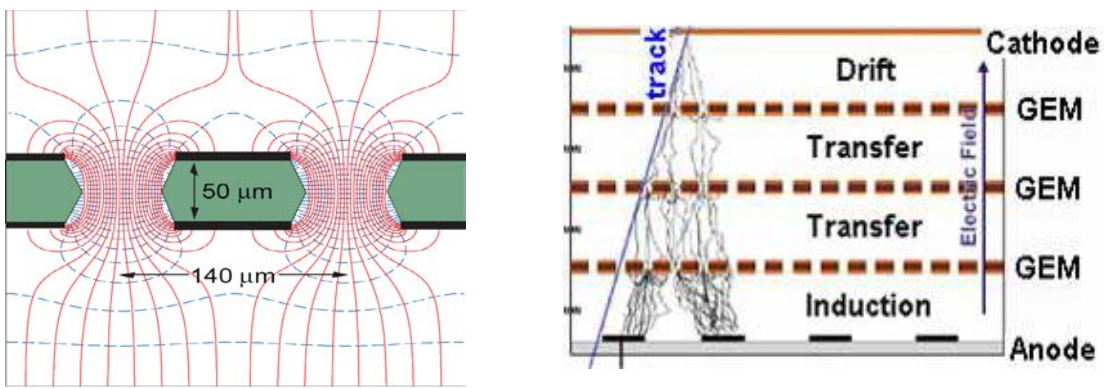

**Fig. 2:** (left) Schematic view and typical dimensions of the hole structure in the GEM amplification cell. Electric field lines (solid) and equipotentials (dashed) are shown16; (right) schematic view of the triple-GEM detector.

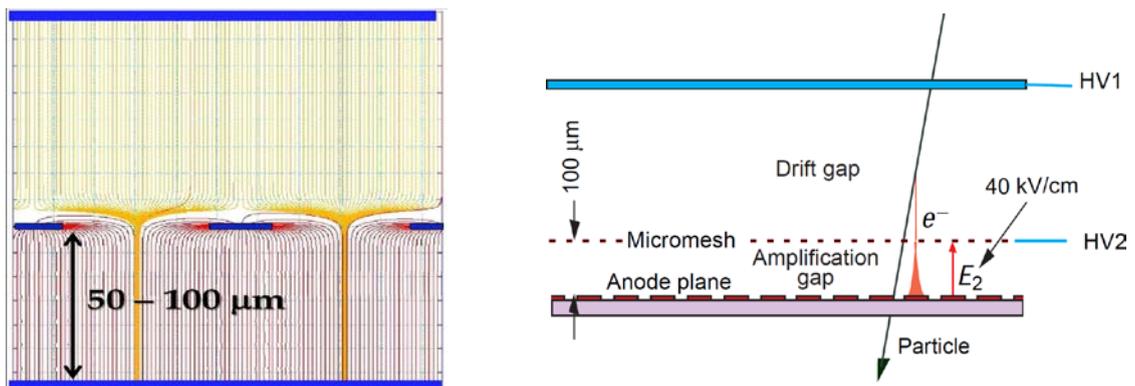

**Fig. 3:** (left) Electric field lines in Micromegas; (rigth) schematic drawing of the Micromegas detector.

Introduced in 1996, a Gas Electron Multiplier (GEM) detector consists of a thin-foil copper-insulator-copper sandwich chemically perforated to obtain a high density of holes. The GEM manufacturing method, developed at CERN, is a refinement of the double-side printed circuit technology. Due to the simultaneous etching from both sides of the foil, the holes acquire a double-conical cross-section. The hole diameter is typically between 25 μm and 150 μm, while the corresponding distance between holes varies between 50 μm and 200 μm. The central insulator is usually (in original design) the polymer kapton, with a thickness of 50 μm. Application of a potential difference between the two sides of the GEM generates the electric field indicated in Fig. 2(left). Each hole acts as an independent proportional counter. Electrons released by the primary ionization particle in the upper drift region (above the GEM foil) are drawn into the holes, where charge multiplication occurs in the high electric field (50–70 kV/cm). Most of avalanche electrons are transferred into the gap below the GEM. Several GEM foils can be placed at short distances (typically 1–2 mm) to distribute the gas amplification among several stages (see Fig. 2(right)). A unique property of GEM detector is a full decoupling of the amplification stage (GEM) and the charge collection electrode - readout board of arbitrary pattern, placed below the last GEM. The signal detected on the PCB is generated only by the avalanche electrons (the ions do not contribute to the signal) and is typically a few tens of nanosecond for 1 mm-wide induction gap.

Introduced in 1996, the Micromegas is a thin parallel-plate avalanche counter, as shown in Fig. 3. It consists of a few mm drift region and a narrow multiplication gap (25–150 μm) between a thin metal grid (micromesh) and the readout electrode (strips or pads of conductor printed on an insulator board). Regularly spaced supports (insulating pillars) guarantee the uniformity of the gap between the anode plane and the micromesh, at the expense of a small, localized loss of efficiency. A low electric field (~ 1 kV/cm) guides the primary ionization electrons in the drift region from their point of creation through the holes of the mesh into the narrow multiplication gap, where they are amplified. The electric field is homogeneous both in the drift and amplification (~ 50–70 kV/cm) regions and only exhibits a funnel-like shape close to the openings of the micromesh: field lines are compressed into a small diameter of the order of few microns, depending on the electric field ratio ($\alpha$) between the multiplication and drift gaps (see Fig. 3(left)). The transverse size of the electron avalanche due to diffusion is of the order of 10–15 μm, depending on the gas mixture, the electric field and the gap width. The anode plane is usually a simple PCB with copper strips or pads of arbitrary shape. The Micromegas retains the rate capability and energy resolution of the parallel-plate counter. The small amplification gap produces a narrow avalanche, giving rise to excellent spatial resolution: 12 μm accuracy, limited by the micromesh pitch, has been achieved for MIPs, and single photo-electron time resolution better than 1 ns. Positive ions are quickly removed by the micromesh; this prevents space-charge accumulation and induces very fast signals (~100 ns length), largely due to ion movement.

The success of GEMs and glass capillary plates triggered in 2003 the development of coarse and more robust structures, "optimized GEM" followed by THGEM gaseous multiplier. These are produced by standard PCB technology: mechanical drilling of 0.3–1 mm diameter holes, etched at their rims to enhance high-voltage stability (see Fig. 4(left)); different PCB materials can be used, of typical thicknesses of 0.4–1 mm and hole spacing of 0.7–1.2 mm. These multipliers exhibit specific features: the electron collection and transport between cascaded elements is more effective than in GEM because the THGEM's hole diameter is larger than the electron's diffusion range when approaching the hole. A novel spark-protected version of thick GEM with resistive electrodes (RETGEM) has been recently developed, where the Cu-clad conductive electrodes are replaced by resistive materials. At low counting rates, the detector operates as a conventional THGEM with metallic electrodes, while in case of discharges the behavior is similar to a resistive-plate chamber.

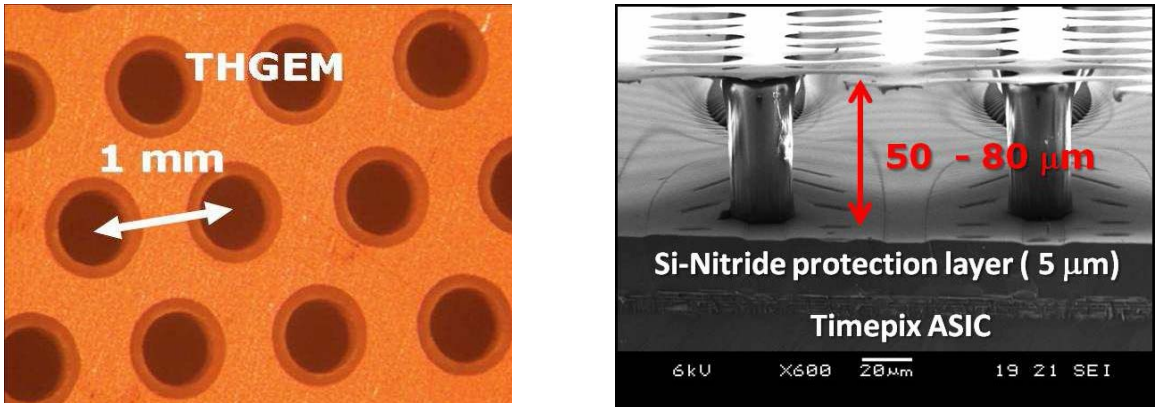

**Fig. 4:** (left) Photo of the Thick GEM (THGEM) multiplier; (right) photo of the "InGrid" detector with ~5 μm silicon-nitride protection layer.

The fine granularity and high-rate capability of GEM and Micromegas devices can be fully exploited by using high-density pixel readout with a size corresponding to the intrinsic width of the detected avalanche charge. While the standard approach for the readout of MPGDs is a segmented strip or pad plane with front-end electronics attached through connectors from the back side, an attractive possibility is to use CMOS pixel chip (without bump-bonded semiconductor sensor) in the gas volume, assembled directly below the GEM or Micromegas amplification structures. These detectors use the input pads of a pixel chip as an integrated charge-collecting anode. With this arrangement signals are induced at the input gate of a charge-sensitive preamplifier (top metal layer of the CMOS chip). Every pixel is then directly connected to the amplification and digitization circuits, integrated in the underlying active layers of the CMOS technology, yielding timing and charge measurements as well as precise spatial information in 3D.

An elegant solution for the construction of Micromegas with pixel readout is the integration of the amplification grid and CMOS chip by means of advanced wafer "post-processing" technology. With this technique, the structure of a thin (1 μm) aluminum grid is fabricated on top of an array of insulating pillars, which stand ~50 μm above the CMOS chip. This novel concept is called "InGrid" (see Fig. 4(right)). The sub-μm precision of the grid dimensions and avalanche gap size results in a uniform gas gain. The grid hole size, pitch, and pattern can be easily adapted to match the geometry of any pixel readout chip. Amongst the most critical items that may affect the long-term operation of the "InGrid" concept is the appearance of destructive sparks across the 50 μm amplification gap, similar to Micromegas. One way to achieve protection is to cover the chip with a thin layer (~ 5 μm) of silicon nitride ($Si_3N_4$) deposited on top of the CMOS ASIC.

## 3. RD51 COLLABOTION

The interest in the technological development and the use of the novel MPGD technologies has led to the establishment in 2008 of the international research collaboration, RD51 at CERN. Nowadays, it involves more than 90 Universities and Research Laboratories from 30 countries in Europe, America, Asia and Africa. All partners are already actively pursuing either basic- or application-oriented R&D involving a variety of MPGD concepts. The RD51 serves as an access point to the MPGD know-how for the world-wide community, being a platform for sharing information, results and experience, and optimizes the cost of R&D developments through sharing of resources, creating common projects and infrastructure.

Many of the MPGD technologies we know today were introduced before RD51 was founded. But with more techniques becoming available (or affordable), new detection concepts are still being introduced and existing ones are substantially improved. Because of a growing interest in the benefits of MPGDs in many fields of research, technologies are being optimized for a wide variety of applications, demonstrating the capabilities of this class of detectors. The RD51 collaboration activities are transversely organized in seven working groups (WG) covering all relevant topics of MPGD-related R&D. A number of tasks are assigned to each WG; for example, detector optimization, discharge protection, ageing and radiation hardness, optimal choice and characterization of gas mixtures and component materials, availability of adequate simulation tools, optimized readout electronics and readout integration with detectors, as well as production and industrialization aspects. Figure 5 lists all WGs and summarizes their objectives and tasks [12].

**Fig. 5:** Organization of the RD51 Collaboration in working groups

Due to their wide variety of geometries and flexible operating parameters, MPGDs are a common choice for tracking, triggering and calorimetry in nuclear- and particle-physics, photon detectors (e.g., RICH), cryogenic detectors for rare events and astroparticle physics (e.g., axion and dark matter searches, neutrino-nucleus scattering, double-beta decay), X-ray and neutron imaging, medical imaging, synchrotron radiation, plasma diagnostics and homeland security applications (e.g., detection of nuclear fission materials or waste in cargo containers by tomography of cosmic ray muons). Many applications area benefit from the MPGD technological advances developed within the framework of the RD51 collaboration activities; however the responsibility for the completion of the application projects lies with the participating institutes themselves. Common themes for applications are low mass, large active areas ($\sim$ m$^2$), high spatial resolution, high-rate capability, and radiation hardness. Today's MPGDs have opened a new era of state-of-the-art technologies and are the benchmarks for the gas detector developments beyond the LHC (see Table 1).

|  | Vertex | Inner Tracker | PID/photo-det. | CALO | MUON Track | MUON Trigger |
|---|---|---|---|---|---|---|
| **ATLAS** | GOSSIP/InGrid | GOSSIP/InGrid |  |  | MM | MM |
| **CMS** |  |  |  | FCAL (GEM,MM) | GEM | GEM |
| **LHCb** |  |  |  |  |  | GEM |
| **ALICE** |  | TPC (GEM) | VHPMID (CsI-THGEM) |  |  |  |
| **Linear Colliders (ILC/CLIC)** |  | TPC (MM,GEM, InGrid) |  | DHCAL (MM,GEM, THGEM) |  |  |

**Table 1:** Examples of ongoing MPGD R&D efforts for the luminosity upgrade of the Large Hadron Collider (sLHC) and the future Linear Collider (ILC or CLIC).

## 4. TECHNOLOGY ADVANCES: LARGE AREA MPGDS

One of the prominent MPGD technology advances is the development of large area GEM, Micromegas and THGEM detectors. Only one decade ago the largest size of MPGDs was around 40 × 40 cm$^2$, limited by existing tools and materials. A big step in the direction of the industrial manufacturing of MPGD's with a unit size of a $\sim$ m$^2$ and spatial resolution, typical of silicon micro-strip devices (30–50 μm), was the development of new fabrication technologies — single mask GEM and "Bulk Micromegas". Applications of large area MPGDs are in high luminosity upgrades of LHC experiments, (semi-) digital hadronic calorimeters (DHCAL, based on a particle flow algorithm (PFA)) for a future LC detectors, light readout of a ring-imaging Cherenkov (RICH) detector and large scale installations that use cosmic ray muons to probe cargo for homeland security applications or temporal tomography densitometric measurements.

Recent developments for fabrication of large area GEM foils are focused on the new technique to overcome the existing limitations: a single-mask technology [13]. The standard technique for creating the GEM hole pattern, involving accurate alignment of two masks, is replaced by a single-mask technology to pattern only the top copper layer. The bottom copper layer is etched after the polyamide, using the holes in the polyamide as a mask (see Fig. 6(left)). A single mask technique overcomes the cumbersome practice of alignment of two masks between top and bottom films within 5–10 μm, which limits the achievable lateral size to ~ 50 cm. This technology generally simplifies the fabrication process to such extent that, especially with large production volumes, the cost per unit of area drops by orders of magnitude. Large area GEMs (1440×700 mm$^2$ foil which is obtained by splicing three 480×700 mm$^2$ foils) can be bent to form cylindrically curved ultra-light tracking systems, without support and cooling structures, as preferred for inner tracker (barrel) and vertex applications [14]. The largest working prototype at present is triple-GEM detector of $1 \times 0.45$ m$^2$ foil size developed for the forward muon system upgrade in the CMS experiment [15].

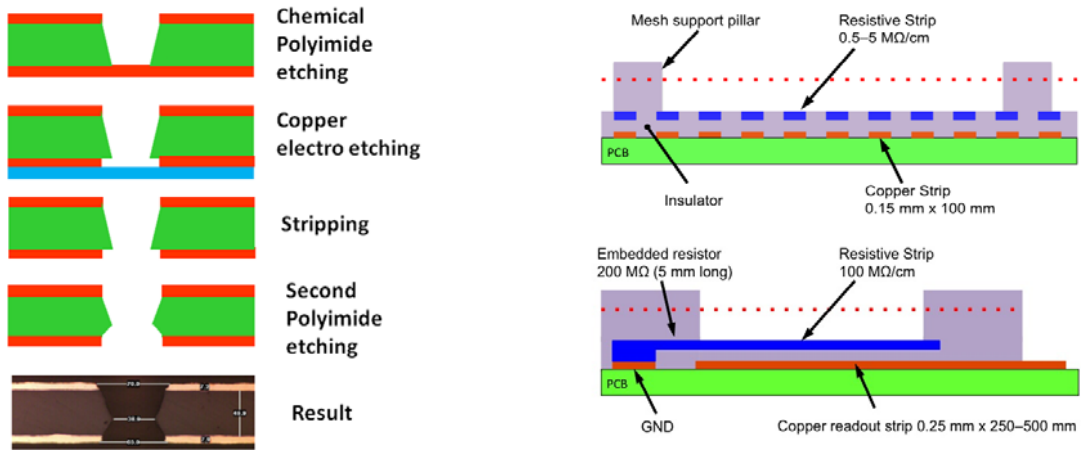

**Fig. 6:** (left) sketch of the single mask GEM technology; (right) sketch of the resistive Micromegas principle, illustrating the resistive protection scheme.

The production of large Micromegas often relies on the "Bulk" Micromegas technique. A photo-imageable coverlay is applied to the PCB with the thickness corresponding to the multiplication gap. A woven mesh and a second layer of photosensitive polyamide film are later laminated on top of the first layer. Exposing the films through appropriate masks create pillars to support the mesh in the active area. Such a "all-in-one" detector, called "Bulk" Micromegas, is robust, shows improved homogeneity, and allows regular production of large, stable and inexpensive detector modules. One of the recent highlights is the development of MPGDs with resistive electrodes for discharge protection. Recently, spark-protected Micromegas was built by adding above the anode readout strips a layer of resistive strips, separated by an insulating layer [16,17]. The readout strips are individually connected to the ground through the large resistor. The principle of the resistive spark protection is schematically shown in Fig. 6(right). It is an appealing solution for tracking in high rate environments (sLHC) or when the spread of avalanche signals onto readout electrodes is desirable (LC-TPC). A rather advanced proponent of large resistive Micromegas is the group working on an upgrade for the ATLAS forward muon spectrometer. The largest detectors built so far and operating smoothly have dimensions of $2.2 \times 0.9$ m$^2$ with the potential to go to even larger module sizes [18]. Many types of resistive readout structures have been developed and validated; either a complete resistive layer over the readout structure or a patterned resistive anode, and often the grounding scheme of these resistive elements is crucial in obtaining optimal performance. Many techniques have been adopted and many more are under development: spraying, screen printing, etching resistive film, filling a negative insulator pattern with resistive paste.

The principle of THGEM-based detectors is well established and recipes to build detectors satisfying all the major requirements have been worked out. Detailed R&D studies dedicated to explore the engineering aspects towards large size detectors are being persuaded in the context of the upgrades of both the COMPASS RICH [19] and the ALICE HMPID [20] systems. No stopping point has been encountered. The COMPASS RICH upgrade has been recently approved - the major issue is represented by the industrial production of large size THGEM boards of the required quality.

For the future Linear Collider (ILC/CLIC) applications, GEM, Micromegas and "InGrid" devices are foreseen as the main options for the Time Projection Chamber (TPC) [21,22]. Spatial resolution reaches a record 50 μm for the TPC applications. The TPC presents a minimum amount of material as required for the best calorimeter and particle-flow performance. Compared to wire chambers, MPGDs offer a number of advantages: negligible E×B track-distortion effects, the narrow Pad Response Function (PRF) and the intrinsic suppression of ion feedback, relaxing the requirement on gating of the devices, and, depending on the design, possibly allowing non-gated operation of the TPC.

Large-area GEM, Micromegas or THGEM are also studied as a potential solution for the highly granular Digital Hadron Calorimeter (DHCAL) [23-25]. To allow the PFA-based approach to calorimetry (3D or imaging calorimeters), the active layers need to have fine longitudinal and transverse segmentation. This allows following charged tracks through the calorimeter with high accuracy, facilitates the separation of hadron showers, and supports the direct measurement of neutral shower energies with adequate resolution. In a sampling DHCAL for the LC, steel absorber plates are alternated with thin detector planes with 1×1cm² MPGD readout elements. Encouraging results have been obtained with ~ m² prototype built from three 33×100cm² single-mask GEM foils and 40×50cm² Micromegas prototypes. Inspired by the work of the ILC/CLIC community, an idea of using PFA calorimetry for the CMS forward calorimetry (both electromagnetic and hadronic) has been recently proposed for the sLHC upgrade (Phase II).

The new development of an integrated Micromegas ("InGrid") on top of silicon micropixel anodes offers a novel and challenging read-out solution. A key element that has to be solved to allow CMOS pixel readout of MPGD for various applications is the production of large area devices. This "InGrid" concept allows to reconstruct tracks and X-ray conversions with unprecedented precision and dE/dx measurement can be performed by electron or even cluster counting. The first prototype of a matrix consisting of $2 \times 4$ Timepix ASICs ("Octopuce") was built in 2010. Work is currently ongoing to design detector consisting of 100 "InGrid" chips; it will use "Octopuce" module as the basic building block for the large-area device [26]. The "InGrid" detector will be used in searching for solar axions in the CAST experiment at CERN and is also being proposed for a TPC at the ILC and for a pixellized tracker (the "gas on slimmed silicon pixels" or "GOSSIP"), aiming for a spatial precision of around 20 μm, for the upgrade of the LHC experiments.

There is a growing demand for cryogenic MPGDs (down to 80 K) for dark matter, rare event searches and neutrino physics. In addition, dark matter experiments require a large target mass and thus the use of liquid argon or xenon. Operation in cryogenic conditions is very challenging: double phase configurations with extraction of primary electrons from the liquid to the gas phase are investigated. In 2011, this concept has been successfully applied to the liquid argon TPC prototype read out by a large area LEM (THGEM) of 76 x 40cm$^2$ size [27]. R&D studies on the possibility to use "GOSSIP" detector for dark matter searches has been also started in the framework of the DARWIN consortium.

## 5. TECHNOLOGY ADVANCES: SCALABLE READOUT SYSTEM (SRS) ELECTRONICS

In parallel to the fabrication of large area MPGD, an important mission of the RD51 collaboration was the development of a portable multi-channel readout DAQ system of scalable architecture. Initiated in 2009, the so-called Scalable Readout System (SRS), has evolved into an open standard supported by many RD51 groups, contributing to SRS hardware, firmware, software and applications. It facilitates the access to an "easy-to-use electronics" with an associated readout, both for small R&D projects and LHC-like experiments, avoiding parallel developments of similar systems.

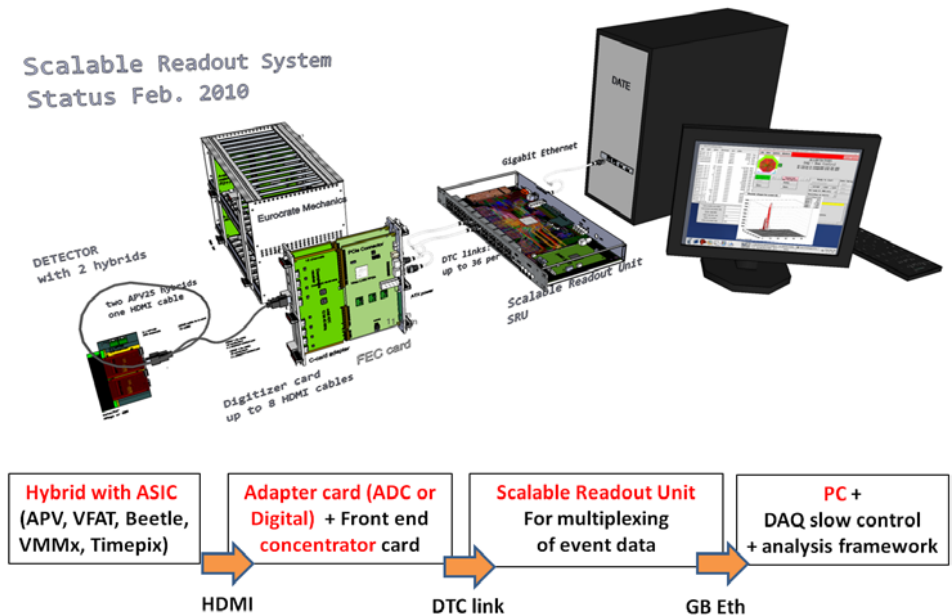

**Fig. 7.** Sketch of the SRS readout architecture, consisting of front-end ASIC on hybrid, a chip-link implemented via HDMI cables, an adapter to a common FEC card and a direct Ethernet connection to an Online PC. Medium-sized systems with several FEC cards require an Ethernet switch and large systems require an Scalable Readout Unit (SRU).

The SRS architecture is composed of a front-end ASIC interfaced to an application-specific adapter card, which digitizes the data stream for FPGA processing on the common front-end concentrator (FEC) card [28-30]. The SRS adapter cards are in principle generic for two types of ASICs: digital or analogue. The Scalable Readout Unit (SRU) multiplexes data for up to 36 FEC, which corresponds to a maximum of 80 k detector channels. All parallel buses in SRS are replaced by point-to-point links and embedded processors are replaced by remote processes. At the end of the chain, a PC with DAQ, slow-control and online software (e.g., DATE) allows to monitor pedestals, noise and analyze physics data (see Fig. 7). A main feature of SRS, apart from its scalability, portability and affordable cost (< 2 EUR/channel), is the free choice of the frontend ASIC (APV, VFAT, Beetle, VMMx, Timepix). More than 40 groups within the RD51 community have already expressed interest in SRS for MPGD applications in basic and applied research; recently, this system has been also used for readout of silicon photo-multipliers.

## 6. TECHNOLOGY ADVANCES: INDUSTRIALIZATION

Nowadays, the CERN Fine Pitch Photolithography Workshop is unique MPGD production facility, where generic R&D, detector components production and quality control take place. To accommodate the development and prototyping (but not necessarily the mass production) of the large area MPGDs, the CERN FinePitch Photolithography workshop has undergone an upgrade in 2010–2012. Approved in 2009 by the CERN management with installation of new equipment being completed in 2012, new MPGD facility will be able to produce $2\times0.5$ m single-mask GEMs, $2\times1$ m "Bulk" Micromegas and $0.8\times0.4$ m THGEM detectors. This opened the road to the very promising R&D on the large-area MPGD-based detectors, in a view of large scale production in industry at a later stage.

A key point that must be solved to further advance cost-effective MPGDs is the manufacturing of large-size detectors and industrialization of the production (technology transfer). RD51 collaboration is representing a reference contact point for firms that are interested in developing competences for MPGD manufacturing. As the technology became more mature for mass production and the technology transfer stabilized, number of industrial companies worldwide expressed strong interest in MPGDs and produced series of prototypes: GEM (TechEtch, Scienergy, New Flex, Techtra), Micromegas (ELVIA, ELTOS, Triangle Lab) and THGEM (Print Electronics, ELTOS). With some of them the contacts have strengthen to the extent that they have signed license agreements and engaged in a technology transfer program defined and coordinated within the RD51 WG6 with the help of CERN Technology Transfer office.

A clear driving force for the qualification of industrial partners is well represented by the large mass production requirements for the upgrade projects of ATLAS, CMS, ALICE and LHCb, planned for the 14TeV energy sLHC upgrades (phase I and II). These needs are listed in the following tables, in case the upgrade projects receive final approval from the LHC collaborations. The aim of RD51 is to achieve the goal of providing qualified industrial production facilities for GEM, Micromegas and THGEM detectors, not only in view of the upgrades for the LHC experiments, but also for the benefit of other physics programs performed worldwide and potential commercial applications that may arise.

**About 1200m² of resistive bulk for small wheel muon stations of Atlas**
**1024 Micromegas layers**

| Sector | Nbr sectors / Nbr chambers/sector / MM layers/chambers | MM layer area (containing rectangle) | Total Nbr MM layers (w/o spares) | Total MM PCB area | Manufacturing plan |
|---|---|---|---|---|---|
| Small | 8x2=16 / 4 / 4x2=8 | From ~0.68m² (696x980) To ~1m² (1420x730) | 512 | 0.88x512 = 450m² | 1st sector 2014 Completion 2016+2017 |
| Large | 8x2=16 / 4 / 4x2=8 | From ~0.96m² (1036X930) To ~1.9m² (2300x835) | 512 | 1.5x512 = 768m² | 1st sector 2014 Completion 2016+2017 |

**About 1000m² of GEM foils for stations 1 and 2 of CMS muon detector**
**216 triple GEM detectors**

| Station | Nbr of modules | Module area (containing rectangle) | Total Nbr of modules (w/o spares) | Total GEM foil area (3ple GEMs) | Manufacturing plan |
|---|---|---|---|---|---|
| GE1/1 | 18x2x2=72 | ~0.43m² (440x990) | 72 | 0.43x72x3 = 93m² | Prototypes 2013+2014 Completion 2016+1017 |
| GE2/1 | 36x2=72 (long) 36x2=72 (short) | ~2.4m² (1251x1911) ~1.6m² (1251x1281) | 144 | (2.4+1.6)x72x3 = 864m² | Prototypes 2013+2014 Completion 2016+1017 |

**About 130m² of GEM foils for Alice TPC upgrade**
**72 triple GEM detectors**

| Module | Nbr of modules | Module area (containing rectangle) | Total Nbr of modules (w/o spares) | Total GEM foil area (3ple GEMs) | Manufacturing plan |
|---|---|---|---|---|---|
| IROC | 18x2=36 | ~0.23m² (460x500) | 36 | 0.23x36x3 =25m² | Yr 2016 |
| OROC | 18x2=36 | ~1m² (880x1120) | 36 | 1x36x3 =108m² | Yr 2016 |

**About 57m² of GEM foils for LHCb muon upgrade**
**144 triple GEM detectors**

| Module | Nbr of modules | Module area | Total Nbr of modules (w/o spares) | Total GEM foil area (3ple GEMs) | Manufacturing plan |
|---|---|---|---|---|---|
| M2-R1 | 48 | ~0.075m² (300x250) | 48 | 0.075x48x3 =11m² | Prototypes 2013+2014 Completion 2015+2016 |
| M2-R2 | 96 | ~0.16m² (600x270) | 96 | 0.16x96x3 =46m² | Prototypes 2013+2014 Completion 2015+2016 |

**Table 2:** Potential needs in large-area MPGD detectors for the ATLAS, CMS, ALICE and LHCb upgrades, if projects are approved by the collaborations. Data are based on the survey performed within the framework of the RD51 collaboration in 2012 [31].

## 7. SUMMARY AND OUTLOOK

Gaseous detectors are fundamental components of all present and planned high energy physics experiments. Advances in photo-lithography and micro-processing techniques during the past decade triggered a major transition in the field from wire chambers to micro-pattern gas-amplification devices. Today, MPGD detectors became a wide-spread tool for the:

- ➢ Future experiments at particle accelerators (e.g., upgrades of muon detectors at the LHC and state-of-the-art concepts for tracking and hadron calorimetry at the future LC);
- ➢ Experiments in nuclear and hadron physics (e.g. KLOE2 at DAFNE, the Panda and CMB experiments at the Facility for Antiproton and Ion Research, STAR at the Relativistic Heavy Ion Collider, SBS at Jefferson Lab and many others);
- ➢ Experiments in astro-particle physics and neutrino physics;
- ➢ Industrial applications such as medical imaging, material science and security inspection.

The interest in the development and use of the novel micro-pattern gas detector technologies has led to the establishment of the research collaboration, RD51 at CERN. The main objective of the R&D program is to advance technological development and application of MPGDs. It is a common platform for sharing of information, results, and experiences, and it supports efforts to make MPGDs suitable for large areas, create common infrastructure at test beams and irradiation facilities, increase cost efficiency, improve ease of use, help with the industrialization of the MPGD technologies, and to develop portable detectors for industrial applications.